\documentstyle[prd,aps,preprint,psfig]{revtex}

\newcommand{\CL}{{\cal L}}

\newcommand{\CO}{{\cal O}}

\newcommand{\bear}{\begin{array}}  \newcommand{\eear}{\end{array}}
\newcommand{\bea}{\begin{eqnarray}}  \newcommand{\eea}{\end{eqnarray}}
\newcommand{\beq}{\begin{equation}}  \newcommand{\eeq}{\end{equation}}
\newcommand{\bef}{\begin{figure}}  \newcommand{\eef}{\end{figure}}
\newcommand{\bec}{\begin{center}}  \newcommand{\eec}{\end{center}}
\newcommand{\non}{\nonumber}  
\newcommand{\lmk}{\left(}  \newcommand{\rmk}{\right)}
\newcommand{\lkk}{\left[}  \newcommand{\rkk}{\right]}
\newcommand{\lhk}{\left \{ }  \newcommand{\rhk}{\right \} }
\newcommand{\lla}{\left \langle }  \newcommand{\rra}{\right \rangle }
\newcommand{\del}{\partial}  
\newcommand{\vect}[1]{\mbox{\boldmath${#1}$}}
\newcommand{\bib}{\bibitem} 
\newcommand{\la}{\left\langle} \newcommand{\ra}{\right\rangle}
 
\newcommand{\gtilde} {~ \raisebox{-1ex}{$\stackrel{\textstyle >}{\sim}$} ~}

\def\IB#1#2#3{{\bf #1}, #2 (19#3)}

\def\IBID#1#2#3{{\it ibid}. {\bf #1}, #2 (19#3)}
\def\IBIDD#1#2#3{{\it ibid}. {\bf #1}, #2 (20#3)}
\def\AA#1#2#3{Astron. Astrophys. {\bf #1}, #2 (19#3)}

\def\APJ#1#2#3{Astrophys. J. {\bf #1}, #2 (19#3)}
\def\APJL#1#2#3{Astrophys. J. Lett. {\bf #1}, L#2 (19#3)}
\def\APJLL#1#2#3{Astrophys. J. Lett. {\bf #1}, L#2 (20#3)}

\def\CQG#1#2#3{Class. Quantum Grav. {\bf #1}, #2 (19#3)}

\def\MNRAS#1#2#3{Mon. Not. R. Astron. Soc. {\bf #1}, #2 (19#3)}
\def\MPLA#1#2#3{Mod. Phys. Lett. A {\bf #1}, #2 (19#3)}
\def\NAT#1#2#3{Nature (London) {\bf #1}, #2 (19#3)}
\def\NATT#1#2#3{Nature (London) {\bf #1}, #2 (20#3)}
\def\NPB#1#2#3{Nucl. Phys. {\bf B#1}, #2 (19#3)}
\def\NPBB#1#2#3{Nucl. Phys. {\bf B#1}, #2 (20#3)}

\def\PLB#1#2#3{Phys. Lett. B {\bf #1}, #2 (19#3)}
\def\PLBB#1#2#3{Phys. Lett. B {\bf #1}, #2 (20#3)}
\def\PLBold#1#2#3{Phys. Lett. {\bf#1B}, #2 (19#3)}

\def\PRD#1#2#3{Phys. Rev. D {\bf #1}, #2 (19#3)}
\def\PRDD#1#2#3{Phys. Rev. D {\bf #1}, #2 (20#3)}

\def\PRL#1#2#3{Phys. Rev. Lett. {\bf#1}, #2 (19#3)}
\def\PRLL#1#2#3{Phys. Rev. Lett. {\bf#1}, #2 (20#3)}

\def\PRT#1#2#3{Phys. Rep. {\bf#1}, #2 (19#3)}
\def\PTP#1#2#3{Prog. Theor. Phys. {\bf #1}, #2 (19#3)}

\begin{document}

\tighten
\draft
\title{Density Fluctuations and Primordial Black Hole Formation
in Natural Double Inflation in Supergravity}
\author{Masahide Yamaguchi}
\address{Research Center for the Early Universe, University of Tokyo,
  Tokyo 113-0033, Japan}

\date{\today}

\maketitle

\begin{abstract}
    We investigate the recently proposed natural double inflation
    model in supergravity. Chaotic inflation first takes place by
    virtue of the Nambu-Goldstone-like shift symmetry. During chaotic
    inflation, an initial value of second inflation (new inflation) is
    set, which is adequately far from the local maximum of the
    potential due to the small linear term in the K\"ahler potential.
    Then, primordial fluctuations within the present horizon scale may
    be produced during both inflations. Primordial fluctuations
    responsible for anisotropies of the cosmic microwave background
    radiation and the large scale structure are produced during
    chaotic inflation, while fluctuations on smaller scales are
    produced during new inflation. Because of the peculiar nature of
    new inflation, they can become as large as $10^{-1}$-$10^{-2}$,
    which may lead to the formation of primordial black holes.
\end{abstract}



\section{Introduction}

\label{sec:int}

Inflation is the most attractive mechanism to generate primordial
density fluctuations responsible for anisotropies of the cosmic
microwave background radiation (CMB) and the large scale structure, in
addition to solving the flatness and the horizon problem
\cite{inflation}.  Realistic inflation models should be constructed in
the context of the supersymmetric theory, especially, its local
version, supergravity (SUGRA) \cite{LR} because supersymmetry (SUSY)
guarantees the flatness of the inflaton against radiative corrections
and gives a natural solution to the hierarchy problem between the
inflationary scale and the electroweak scale \cite{SUSY}.

New inflation \cite{newinf} is very attractive in the context of SUGRA
because it takes place at a low energy scale and naturally leads to a
sufficiently low reheating temperature to avoid the overproduction of
gravitinos. However, roughly speaking, new inflation has two severe
problems \cite{inflation}. One is the initial value problem: that is,
the inflaton must be fine-tuned near the local maximum of the
potential for sufficient inflation. The other is the flatness
(longevity) problem: that is, why the universe lives so long beyond
the Planck time. Asaka, Kawasaki, and the present author \cite{AKY}
found that, due to gravitationally suppressed interactions with
particles in the thermal bath, the inflaton can dynamically go to the
local maximum of its potential. However, the other problem still
exists unless the universe is open at the beginning. Izawa, Kawasaki,
and Yanagida \cite{IKY} considered another type of inflation (called
preinflation) which takes place before new inflation and drives the
inflaton for new inflation dynamically toward the local maximum of its
potential. If preinflation is chaotic inflation \cite{chaoinf}, the
longevity problem is solved too.

It, however, is believed to be difficult to realize chaotic inflation
in SUGRA. This is mainly because a scalar potential in minimal SUGRA
has an exponential factor with the form $\exp(|\phi|^2/M_G^2)$ so that
any scalar field $\phi$ cannot take a value much larger than the
reduced Planck scale $M_G\simeq 2.4\times 10^{18}$ GeV. Several
supergravity chaotic inflation models were proposed by use of
functional degrees of freedom of the K\"ahler potential in SUGRA
\cite{GL,MSY2}. But, there are no symmetry reasons to have such
proposed forms. Recently, Kawasaki, Yanagida, and the present author
\cite{KYY} proposed a natural model of chaotic inflation in SUGRA by
use of the Nambu-Goldstone-like shift symmetry. Motivated by this
model, Yokoyama and the present author \cite{YY} considered chaotic
inflation followed by new inflation, where chaotic inflation first
takes place around the Planck scale to solve the longevity problem and
gives an adequate initial condition for new inflation. Thus double
inflation has been proposed as a solution of initial condition
problems of some types of inflation \cite{IKY,doublehyb}. (See Refs.
\cite{inithyb,doublehyb} for the initial value problem and its
solution of hybrid inflation.)

On the other hand, double inflation is also motivated by observational
results. One motivation is to reconcile predicted spectra with
observations of the large scale structure \cite{double}. It is known
that a standard cold dark matter (CDM) model in a flat universe with a
nearly scale-invariant spectrum cannot reproduce the observation of
the large scale structure. Furthermore, the recent observations of
anisotropies of the CMB by the BOOMERANG experiment \cite{BOOMERANG}
and MAXIMA experiment \cite{MAXIMA} found a relatively low second
acoustic peak. Another motivation is to produce primordial black holes
(PBHs) \cite{PBH,PBH2,PBH3}. Massive compact halo objects (MACHOs) are
observed through gravitational microlensing effects \cite{MACHO},
which are a possible candidate of dark matter.  Furthermore, PBHs
evaporating now may be a source of antiproton flux observed by the
BESS experiment \cite{BESS} or responsible for short gamma ray bursts
(GRBs) \cite{gamma}. Though many double inflations have been
considered, they are often discussed in a simple toy model with two
massive scalar fields. However, a natural double inflation model in
SUGRA is recently proposed \cite{YAMA}, where there are no initial
condition problems and the model parameters are natural in the 't
Hooft sense \cite{tHooft}. In this model, chaotic inflation takes
place first of all, during which an initial value of new inflation is
dynamically set due to the supergravity effects. It can be adequately
far from the local maximum of the potential due to the small linear
term of the inflaton in the K\"ahler potential. Therefore primordial
density fluctuations responsible for the observable universe can be
attributed to both inflations, that is, chaotic inflation produces
primordial fluctuations on large cosmological scales and new inflation
on smaller scales.\footnote{In Ref. \cite{YY} the initial value of new
inflation is so close to the local maximum of the potential for new
inflation that the universe enters a self-regenerating stage
\cite{eternal,sr}. Therefore primordial fluctuations responsible for
the observable universe are produced only during new inflation.
Furthermore, even if chaotic inflation proposed in Ref. \cite{KYY} is
adopted as preinflation in Ref. \cite{IKY}, the same situation occurs,
that is, second inflation becomes eternal inflation because the
superpotential in Ref. \cite{KYY} vanishes during chaotic inflation.}
The energy scale of new inflation becomes of the same order as the
initial value of new inflation so that produced density fluctuations
may become as large as the order of unity due to the peculiar nature
of new inflation, which straightforwardly may lead to PBHs formation.

In this paper we minutely investigate the recently proposed natural
double inflation model in supergravity, especially, primordial density
fluctuations produced during inflation. Then, the PBHs formation is
discussed.

\section{Model and dynamics}

\subsection{Model}

In this section we briefly review the double inflation model in
supergravity proposed recently \cite{YAMA}. We introduce an inflaton
chiral superfield $\Phi(x,\theta)$ and assume that the model,
especially, K\"ahler potential $K(\Phi,\Phi^{\ast},\dots)$ is a
function of $\Phi+\Phi^{\ast}$, which enables the imaginary part of
the scalar component of the superfield $\Phi$ to take a value larger
than the gravitational scale, and leads to chaotic inflation. Such a
functional dependence of $K$ can be attributed to the
Nambu-Goldstone-like symmetry introduced in Ref. \cite{KYY}. We also
introduce a spurion superfield $\Xi$ describing the breaking of the
shift symmetry and extend the shift symmetry as follows,
\bea
  \Phi &\rightarrow& \Phi + i~C M_{G}, \non \\
  \Xi  &\rightarrow& \lmk\frac{\Phi}{\Phi + i~C M_{G}}\rmk^{2} \Xi,
  \label{eq:shift}
\eea
where $C$ is a dimensionless real constant. Below, the reduced Planck
scale $M_{G}$ is set to be unity. Under this shift symmetry, the
combination $\Xi\Phi^{2}$ is invariant. Inserting the vacuum value
into the spurion field, $\la \Xi \ra = \lambda$, softly breaks the
above shift symmetry. Here, the parameter $\lambda$ is fixed with a
value much smaller than unity representing the magnitude of breaking
of the shift symmetry (\ref{eq:shift}).

We further assume that in addition to the shift symmetry, the
superpotential is invariant under the U$(1)_{R}$ symmetry because it
prohibits a constant term in the superpotential. The above K\"ahler
potential is invariant only if the $R$-charge of $\Phi$ is zero. Then,
we are compelled to introduce another supermultiplet $X(x,\theta)$
with its $R$-charge equal to two, which allows the linear term $X$ in
the superpotential. As shown in Ref. \cite{YAMA}, for successful
inflation, the absolute magnitude of the coefficient of the linear
term $X$ must be at most of the order of $|\lambda|$, which is much
smaller than unity. Therefore in order to suppress the linear term of
$X$ in the superpotential, we introduce the $Z_{2}$ symmetry and a
spurion field $\Pi$ with odd charge under the $Z_{2}$ symmetry and
zero $R$-charge. The vacuum value $\la \Pi \ra = v$ softly breaks the
$Z_{2}$ symmetry and suppress the linear term of $X$.  Then, the
general superpotential invariant under the shift, U$(1)_{R}$ and
$Z_{2}$ symmetries is given by
\bea
  W = \alpha_{1} \Pi X \lhk 1 + \alpha_{2}(\Xi\Phi^{2})^{2} 
                               + \cdots \rhk
      - \alpha_{3} X \lhk \Xi\Phi^{2} + \alpha_{4}(\Xi\Phi^{2})^{3} 
                               + \cdots
      \rhk,
  \label{eq:superpotential}
\eea
where we have assumed the $R$-charge of $\Xi$ vanish and $\Phi$, $X$,
and $\Xi$ fields are odd under the $Z_{2}$ symmetry.\footnote{The
oddness of the spurion field $\Xi$ under the $Z_{2}$ symmetry implies
that it breaks both the shift symmetry and the $Z_{2}$ symmetry at
once. So, we expect that the magnitudes of the breaking of both the
$Z_{2}$ and the shift symmetries are of the same order. We hope that
the yet unknown mechanism simultaneously gives the spurion field $\Xi$
and $\Pi$ the vacuum values and such a mechanism be realized, for
example, in the superstring theory.}(See table I in which charges for
superfields are shown.) Here, $\alpha_{i}$ are complex constants of
the order of unity.

After inserting vacuum values of spurion fields $\Xi$ and $\Pi$, the
superpotential is given by
\bea
  W = v X \lhk 1 + \alpha_{2}(\lambda \Phi^{2})^{2} 
                               + \cdots \rhk
      - X \lhk \lambda \Phi^{2} + \alpha_{4}(\lambda \Phi^{2})^{3} 
                               + \cdots
      \rhk.
\eea
Here, the complex constants $\alpha_{1}$ and $\alpha_{3}$ are
renormalized into $v$ and $\lambda$. Though the above superpotential
is not invariant under the shift and the $Z_{2}$ symmetries, the model
is completely natural in the 't Hooft's sense \cite{tHooft} because we
have enhanced symmetries in the limit $\lambda$ and $v \rightarrow 0$.
As long as $|\Phi| \ll |\lambda|^{-1/2}$, higher order terms with
$\alpha_{i}$ of the order of unity become irrelevant for the dynamics.
Therefore we can safely omit them in the following discussion. After
all, we use, in the following analysis, the superpotential,
\bea
  W &\simeq& vX - \lambda X\Phi^{2} \\
    &=& vX(1 - g\Phi^{2}),
\eea
with $g \equiv \lambda / v$. Though, generally speaking, only a
constant can become real by the use of the phase rotation of the $X$
field, below we set both constants $v$ and $\lambda$ ($g$) to be real
for simplicity.\footnote{The dynamics of the general case is discussed
in Ref. \cite{YAMA}.}

The K\"ahler potential neglecting a constant term and higher order
terms is given by
\beq
  K = v_{2}(\Phi + \Phi^{\ast}) 
     + \frac12 (\Phi + \Phi^{\ast})^{2} 
     + XX^{\ast}.
  \label{eq:kahler}
\eeq
Here $v_{2} \sim v$ is a real constant representing the breaking
effect of the $Z_{2}$ symmetry. Here and hereafter, we use the same
characters for scalar with those for corresponding supermultiplets.

\subsection{Dynamics}

Now that the K\"ahler potential and the superpotential are specified,
the Lagrangian density $L(\Phi,X)$ for the scalar fields $\Phi$ and
$X$ is given by
\beq
  L(\Phi,X) = \partial_{\mu}\Phi\partial^{\mu}\Phi^{\ast} 
  + \partial_{\mu}X\partial^{\mu}X^{\ast}
         -V(\Phi,X).
\eeq
The scalar potential $V$ of the chiral superfields $X(x,\theta)$ and
$\Phi(x,\theta)$ in supergravity is given by
\beq
  V = v^{2} e^{K} \lkk
      \left|1 - g\Phi^{2}\right|^{2}(1-|X|^{2}+|X|^{4}) 
       + |X|^{2} \left
         |-2g\Phi + (v_{2}+\Phi+\Phi^{\ast})(1-g\Phi^{2})
                 \right|^{2}
                  \rkk.
\eeq
Now, we decompose the scalar field $\Phi$ into real and imaginary
components,
\beq
  \Phi = \frac{1}{\sqrt{2}} (\varphi + i \chi).
\eeq
Then, the Lagrangian density $L(\varphi,\chi,X)$ is given by
\beq
  L(\varphi,\chi,X) = 
              \frac{1}{2}\partial_{\mu}\varphi\partial^{\mu}\varphi 
              + \frac{1}{2}\partial_{\mu}\chi\partial^{\mu}\chi 
              + \partial_{\mu}X\partial^{\mu}X^{*}
              -V(\varphi,\chi,X),
\eeq
with the potential $V(\varphi,\chi,X)$ given by
\bea \hspace{-1.0cm}
  V(\varphi,\chi,X)
    &=& v^{2} e^{-\frac{v_{2}^{2}}{2}}
           \exp \lhk \lmk \varphi + \frac{v_{2}}{\sqrt{2}} \rmk^{2} 
                     + |X|^{2}
                \rhk \non \\ 
    && \hspace{0.0cm} \times
         \lhk~\lkk 
              1 - g (\varphi^{2} - \chi^{2}) 
             + \frac14~g^{2} (\varphi^{2} + \chi^{2})^{2}
              \rkk
             (1-|X|^{2}+|X|^{4}) 
         \right. \non \\ 
    && \hspace{0.5cm}
             +~|X|^{2} 
              \lkk~
                2g^{2}(\varphi^{2}+\chi^{2})
              \right. \non \\
    && \hspace{2.0cm}
                - (v_{2}+\sqrt{2}\varphi) \lhk
                  \sqrt{2}~g\varphi
                    \lkk~2 - g (\varphi^{2} - \chi^{2}) 
                    ~\rkk - 2\sqrt{2}~g^{2}\varphi\chi^{2}
                    \rhk \non \\
    && \hspace{2.0cm} \left. \left.    
                 +(v_{2}+\sqrt{2}\varphi)^{2} 
                    \lhk
                      1 - g (\varphi^{2} - \chi^{2}) 
                    + \frac14~g^{2} 
                    ~(\varphi^{2} + \chi^{2})^{2}
                    \rhk
               ~\rkk
                      ~\rhk.
\eea

Because of the exponential factor, $\varphi$ and $X$ rapidly goes down to
$\CO(1)$. On the other hand, $\chi$ can take a value much larger than
unity without costing exponentially large potential energy. Then the
scalar potential is approximated as
\beq
  V \simeq \lambda^{2}
                   \lmk \frac{\chi^{4}}{4}      
                        + 2 \chi^{2} |X|^{2}
                   \rmk,
  \label{eq:twopot}
\eeq
with $\lambda = g v$. Thus the term proportional to $\chi^{4}$ becomes
dominant and chaotic inflation can take place. Then, using the
slow-roll approximation, the $e$-fold number $\widetilde{N}$ during
chaotic inflation is given by
\beq
  \widetilde{N} \simeq \frac{\chi_{\widetilde{N}}^{2}}{8}.
  \label{eq:chaoe}
\eeq
The effective mass squared of $\varphi$, $m_{\varphi}^{2}$, during
chaotic inflation becomes
\beq
  m_{\varphi}^{2} \simeq \frac{\lambda^{2}}{2} \chi^{4}
                  \simeq 6H^{2} \gg \frac94 H^{2},
                  ~~~~~~H^{2} \simeq \frac{\lambda^{2}}{12}\chi^{4},
\eeq
where $H$ is the hubble parameter at that time. Therefore $\varphi$
oscillates rapidly around the minimum $\varphi_{min}$ so that its
amplitude damps in proportion to $a^{-3/2}$ with $a$ being the scale
factor. Here, the potential minimum for $\varphi$, $\varphi_{min}$,
during chaotic inflation is given by
\beq
  \varphi_{min} \simeq - v_{2}/\sqrt{2}.
\eeq
Thus the initial value of the inflaton $\varphi$ of second inflation
(new inflation) is set dynamically during chaotic inflation.

On the other hand, the mass squared of $X$, $m_{X}^{2}$, is dominated
by
\beq
  m_{X}^{2} \simeq 2 \lambda^{2} \chi^{2} \simeq \frac{24}{\chi^2}H^2,   
  \label{mhratio}
\eeq
which is much smaller than the hubble parameter squared until
$\chi^{2} \sim 24$ so that $X$ also slow-rolls. In order to analyze
the dynamics of $X$, we set $X$ to be real and positive making use of
the freedom of the phase choice. In this regime classical equations of
motion for $X$ and $\chi$ are given by
\bea
  3H &\dot{X}& \simeq - m_{X}^{2} X,  \label{Xeq} \\
  3H &\dot{\chi}& \simeq - \lambda^{2} \chi^{3}, \label{chieq}
\eea
which leads to
\beq
  \lmk \frac{X}{X_{i}} \rmk \simeq 
    \lmk \frac{\chi}{\chi_{i}} \rmk^{2},   \label{propto}
\eeq
where $X_{i}$ and $\chi_{i}$ are the initial values of $X$ and $\chi$
fields. This relation holds actually if and only if quantum
fluctuations are unimportant for both $\chi$ and $X$. First of all,
for $\chi$, the comparison of the magnitude of quantum fluctuations
and that of the classical evolution during one hubble time shows that
quantum fluctuations become dominant if $\chi \gtrsim \lambda^{-1/6}$,
when the universe enters the self-reproduction stage of eternal
inflation \cite{eternal,sr}. So, we consider only the regime with
$\chi \ll \lambda^{-1/6}$, where the classical equation of motion Eq.
(\ref{chieq}) is valid. Next, we estimate the amplitude of quantum
fluctuations for $X$, Using the Fokker-Planck equation for the
statistical distribution function of $X$ based on the stochastic
inflation method of Starobinsky \cite{stochastic}, the
root-mean-square (rms) of quantum fluctuations for $X$, $\lla
\lmk\Delta X\rmk^2 \rra$, is given by \cite{YY}
\beq
   \sqrt{\lla \lmk\Delta X\rmk^2 \rra} \simeq
   \frac{\lambda^{1/3}}{8\pi\sqrt{6}}~\chi^2.  \label{Xyuragi}
\eeq
On the other hand, using $X_{i} = \CO(1)$ and $\chi_{i} \sim
\lambda^{-1/6}$, the classical value of $X$ becomes $X \sim
\lambda^{1/3} \chi^{2}$. Thus, since $\lambda \ll 1$, the amplitude of
$X$ becomes much smaller than unity by the time $\chi \simeq
\sqrt{24}$, when the effective mass squared $m_{X}^{2}$ is comparable
with $H^{2}$. Thereafter, $X$ rapidly oscillates around the origin and
its amplitude damps in proportion to $a^{-3/2}$ even more. Thus our
approximation that both $\varphi$ and $X$ are much smaller than unity
is consistent throughout the chaotic inflation regime.

As $\chi$ becomes of order of unity, either the constant term $v^{2}$
or the term with $v^{2} g \chi^{2}$ becomes dominant. In the former
case, small hybrid inflation takes place, which is followed by new
inflation. Hence there is no break between chaotic and new inflation.
On the other hand, in the latter case, $\chi$ rapidly oscillates
around the origin until new inflation starts so that there is a break
between them, though the scale factor grows twice at most. A little
numerical calculation shows that if $g \gtilde 1.1$, we have a break
between chaotic and new inflation.

Next let us investigate when new inflation starts. The potential with
$X \simeq 0$ is approximated as
\bea
  V(\varphi,\chi,X \simeq 0)
    &\simeq& v^{2} e^{-\frac{v_{2}^{2}}{2}}
           \exp \lmk \varphi + \frac{v_{2}}{\sqrt{2}} 
                \rmk^{2} 
                \non \\ 
    && \hspace{-2.0cm} \times
         \lkk 
             \lmk 
               1 - \frac{g}{2} \varphi^{2} 
             \rmk^{2}
            + \chi^{2}
             \lmk
               g + \frac{g^{2}}{2} \varphi^{2}
                   + \frac{g^{2}}{2} \chi^{2}
             \rmk
         \rkk.  
\eea
The global minima are given by $\varphi^{2} = 2/g$ and $\chi = 0$. The
mass squared for $\varphi$, $m_{\varphi}^{2}$, reads
\beq
  m_{\varphi}^{2} 
      \simeq - (g - 1) + (g + \frac12 g^{2}) \chi^{2}.
\eeq
Thus new inflation begins when $\chi \simeq \chi_{crit}$ given by
\beq
  \chi_{crit} =  \frac{2}{g} 
                        \sqrt{\frac{g - 1}{g + 2}}.
\eeq

Once new inflation begins, $\chi$ rapidly goes to zero because the
effective mass squared becomes $m_{\chi}^{2} \simeq 6 g H^{2} \ge 6
H^{2}$. Then, for $\chi \simeq 0$ and $X \ll 1$, the potential is
given by
\beq
  V(\varphi,\chi \simeq 0,X \ll 1) \sim
    v^{2} \lhk 1 - (g - 1) \widetilde\varphi^{2} 
                 + 2 (g - 1)^{2} \widetilde\varphi^{2}|X|^{2}
                 + \cdots
          \rhk,
\eeq
where $\widetilde\varphi = \varphi - \varphi_{max}$ and $\varphi_{max}
\equiv v_{2}/[\sqrt{2}(g-1)]$. Thus if $g \ge 1$ ($\lambda \ge v$),
new inflation takes place and $\varphi$ rolls down slowly toward the
vacuum expectation value $\eta = \sqrt{2/g}$.

Before new inflation starts, $\varphi$ stays at $\varphi_{min}$, which
is different from $\varphi_{max}$. Then, the initial value of
$\widetilde\varphi$, $\widetilde\varphi_{i}$ for new inflation is given by
\beq
  \widetilde\varphi_{i} = - \frac{v_{2}}{\sqrt{2}}
                         \frac{g}{g-1}.
\eeq
On the other hand, the amplitude of quantum fluctuations of $\varphi$
is estimated as $\delta\varphi_{q} \sim v / (2\pi\sqrt{3})$. Using the
fact that $g \ge 1$, we find that quantum fluctuations do not dominate
the dynamics unless $v_{2} \ll v$.

The total $e$-folding number $N_{new}$ during new inflation is given
by
\beq
  N_{new} \simeq \frac{1}{2(g-1)}
                \ln \left| \frac{\sqrt{2}}{v_{2}}
                           \frac{g-1}{g} \right|.
\eeq
Then the total $e$-folding number $N$ is given by $N = \widetilde{N} +
N_{new}$. We set $N_{\rm COBE} = 60$ for simplicity, when the physical
wavenumber of the mode ($k_{\rm COBE}$) corresponding to the Cosmic
Background Explorer~(COBE) scale exits the horizon, that is, $k_{\rm
COBE}/(a_{(N=60)} H_{(N=60)}) = 1$.

In case $N_{new} \gtrsim 60$, primordial density fluctuations
responsible for the observable universe are produced only during new
inflation. Otherwise, chaotic inflation produces primordial
fluctuations on large cosmological scales and new inflation on smaller
scales. In this paper we consider only the latter case.

After new inflation, $\varphi$ oscillates around the global minimum
$\eta$ and the universe is dominated by a coherent scalar-field
oscillation of $\sigma \equiv \varphi -\eta$. Expanding the
exponential factor $e^{v_{2}\varphi + \varphi^2} $ in $e^K$,
\beq
  e^{v_{2}\varphi + \varphi^2} = e^{\eta^2}(1+2\eta\sigma+\cdots ),
\eeq
we find that $\sigma$ has gravitationally suppressed linear
interactions with all scalar and spinor fields including minimal
supersymmetric standard model (MSSM) particles. For example, let us
consider the Yukawa superpotential $W = y_{i}D_{i}HS_{i}$ in MSSM,
where $D_{i}$ and $S_{i}$ are doublet (singlet) superfields, $H$ is a
Higgs superfield, and $y_{i}$ is a Yukawa coupling constant. Then the
interaction Lagrangian is given by
\beq
  \CL_{\rm int} \sim 
     y_{i}^{2} \eta \sigma D_{i}^{2} S_{i}^{2} + \cdots,
\eeq
which leads to the decay width $\Gamma$ given by 
\beq
  \Gamma \sim \sum_{i} y_{i}^4 \eta^2 m_{\sigma}^3.
  \label{eq:gamma}
\eeq
Here $m_{\sigma} \simeq 2\sqrt{g_{R}}e^{\sqrt{2/g_{R}}}v$ is the mass
of $\sigma$. Thus the reheating temperature $T_{R}$ is given by
\beq
  T_{R} \sim 0.1 \bar{y} \eta m_{\varphi}^{3/2},
\eeq
where $\bar{y}=\sqrt{\sum_{i} y_{i}^4}$. Taking
$\bar{y} \sim 1$, the reheating temperature $T_{R}$ is given by
\beq
  T_{R} \sim v^{3/2} \lesssim \lambda^{3/2}. 
\eeq
As shown later, the upper bound of $\lambda$ is given by $\lambda <
1.2 \times 10^{-6}$. Hence the reheating temperature $T_{R}$ is
constrained as
\beq
  T_{R} \lesssim 10^{-9} \sim 10^{9}~{\rm GeV},
\eeq
which is low enough to avoid the overproduction of gravitinos in a
wide range of the gravitino mass \cite{Ellis}.

\section{Density fluctuations and PBHs formation}

\subsection{Density fluctuations}

In this section we investigate primordial density fluctuations
produced by this double inflation model. First of all we consider
density fluctuations produced during chaotic inflation. As shown in
the previous section, there are two effectively massless fields,
$\chi$ and $X$, during chaotic inflation. Using Eq. (\ref{eq:twopot})
and adequate approximations, the metric perturbation in the
longitudinal gauge $\Phi_{A}$ can be estimated as \cite{PS}
\bea
  \Phi_{A} &=& - \frac{\dot{H}}{H^{2}} C_{1}
               - 16 \frac{X^{2}}{\chi^{2}} C_{2}, \non \\
  C_{1} &=& H \frac{\delta\chi}{\dot{\chi}}, \non \\
  C_{2} &=& H \lmk \frac{\delta\chi}{\dot{\chi}} 
                  - \frac{\delta X}{\dot{X}}
              \rmk 
            \frac{2}{\chi^{2}},
\eea
where the dot represents time derivative, the term with $C_{1}$
corresponds to the growing adiabatic mode, and the term with $C_{2}$
the nondecaying isocurvature mode. You should notice that only $\chi$
contributes to growing adiabatic fluctuations. Then, with the fact
that $X \ll 1$, the amplitude of curvature perturbation $\Phi_{A}$ on
the comoving horizon scale at $\chi=\chi_{\widetilde{N}}$ is
given by the standard one-field formula and reads
\beq
  \Phi_{A}(\widetilde{N})
            \simeq  \frac{f}{2\sqrt{3}\pi}
                  \frac{\lambda\chi_{\widetilde{N}}^{3}}{8}, 
  \label{eq:gpotentialc}
\eeq
where $f=3/5~(2/3)$ in the matter (radiation) domination. If $N_{new}
\lesssim 60$, the comoving scale corresponding to the COBE scale exits
the horizon during chaotic inflation. Defining $\widetilde{N}_{\rm
COBE}$ as the $e$-folding number during chaotic inflation,
corresponding to the COBE scale, the COBE normalization requires
$\Phi_{A}(\widetilde{N}_{\rm COBE}) \simeq 3\times 10^{-5}$
\cite{COBE}. Then the scale $\lambda$ is given by
\beq
    \lambda \simeq 4.2 \times 10^{-3} 
        \chi_{\widetilde{N}_{\rm COBE}}^{-3}. 
    \label{eq:COBEnor}
\eeq
The spectral index $n_{s}$ is given by
\beq
  n_{s} \simeq 1 - \frac{3}{\widetilde{N}_{\rm COBE}}.
\eeq
Since the COBE data shows $n_s = 1.0 \pm 0.2$ \cite{COBE},
$\widetilde{N}_{\rm COBE} \ge 15$, which leads to $\lambda < 1.2
\times 10^{-6}$.

Next let us discuss density fluctuations produced during new
inflation. In this case, both $\widetilde\varphi$ and $X$ are effectively
massless fields. As with the case of chaotic inflation, the metric
perturbation in the longitudinal gauge $\Phi_{A}$ can be estimated as
\cite{PS},
\bea
  \Phi_{A} &=& - \frac{\dot{H}}{H^{2}} C_{1}
               - 4 (g - 1)^{3} X^{2}\widetilde\varphi^{2} C_{2}, \non \\
  C_{1} &=& H \frac{\delta\widetilde\varphi}{\dot{\widetilde\varphi}}, \non \\
  C_{2} &=& H \lmk \frac{\delta\widetilde\varphi}{\dot{\widetilde\varphi}} 
                  - \frac{\delta X}{\dot{X}}
              \rmk 
                2 (g - 1)\widetilde\varphi^{2}.
\eea
In this case, also, only $\widetilde\varphi$ contributes to growing
adiabatic fluctuations.  Then, with the fact that $X \ll 1$, the
amplitude of curvature perturbation $\Phi_{A}$ on the comoving horizon
scale at $\widetilde\varphi=\widetilde\varphi_{N}$ is given by
\beq
  \Phi_{A}(N) \simeq \frac{f}{2\sqrt{3}\pi} 
       \frac{v}{ 2 (g - 1) \widetilde\varphi_{N}}.
  \label{eq:gpotentialn}
\eeq
The spectral index $n_s$ of the density fluctuations is given by
\beq
    n_s \simeq 1 - 4 (g - 1).
\eeq

Here we relate the $e$-folding number $N < N_{new}$ with the comoving
wave number $k$. Notice that the hubble parameter during new inflation
is much smaller than that during the early stage of chaotic inflation
[$\widetilde{N} = \CO(10)$].  Then, the $e$-folding number $N <
N_{new}$ when the comoving wave number $k = k_{\rm COBE}~e^{(60-N')}$
($N_{\rm COBE}=60$) exits the horizon during new inflation is
determined by
\beq
  \frac{k_{\rm COBE}~e^{(60-N')}}{a_{(N=60)}~e^{(60-N)}~H_{N}} = 1.
\eeq 
Using $H_{N} \simeq v / \sqrt{3}$, $H_{(N=60)} \simeq \lambda
\chi_{\widetilde{N}_{\rm COBE}}^{2}/(2\sqrt{3})$, and Eq.
(\ref{eq:chaoe}), the correspondence is given by
\bea
  N &=& N' + \ln \lmk \frac{2}{g \chi_{\widetilde{N}_{\rm COBE}}^{2}} \rmk,
        \non \\
    &=& N' - \ln 4 g \widetilde{N}_{\rm COBE}.
\eea
The deviation from the standard correspondence ($N = N'$) is not
negligible for $\widetilde{N}_{\rm COBE} = \CO(10)$.\footnote{The
hubble parameter during chaotic inflation changes significantly ($H
\propto \widetilde{N}$). Hence, the deviation from the standard
correspondence may be also significant during chaotic inflation. The
$e$-folding number $\widetilde{N} < \widetilde{N}_{\rm COBE}$ when the
comoving wave number $k = k_{\rm COBE} e^{(\widetilde{N}_{\rm
COBE}-N')}$ exits the horizon during chaotic inflation is given by $N'
= \widetilde{N} + \ln (\widetilde{N}_{\rm COBE}/\widetilde{N})$.}

You should also notice that $\widetilde\varphi \sim v_{2}$ at the
beginning of new inflation. Then, since $v \sim v_{2}$, the amplitude
of curvature perturbation $\Phi_{A}$ can become as large as the order
unity, which may lead to PBHs formation.

Before discussing PBHs formation, let us comment on the case with a
break between chaotic and new inflation. As shown before, we have such
a break if $g \gtrsim 1.1$. In this case some of the comoving wave
numbers which exit the horizon during chaotic inflation reenter the
horizon and again exit during new inflation. Therefore, for such
modes, we need to compare the amplitude of quantum fluctuations
induced during chaotic and new inflation. Following the procedure as
done in Ref. \cite{PBH3}, we can easily show that fluctuations induced
during chaotic inflation are a little less than newly induced
fluctuations during new inflation. Thus we conclude that the
fluctuations of $\varphi$ induced in chaotic inflation can be
neglected when we estimate density fluctuations produced during new
inflation.

\subsection{Primordial black holes formation}

PBHs have been paid renewed attention to because they may explain the
existence of massive compact halo objects (MACHOs) \cite{MACHO} and
become a part of cold dark matter. Furthermore, PBHs are responsible
for antiproton fluxes observed by the BESS experiments \cite{BESS} or
short gamma ray bursts \cite{gamma}.

Carr and Hawking first discussed PBHs formation and showed that in the
radiation dominated universe, a black hole is formed soon after the
perturbed region reenters the horizon if the amplitude of density
fluctuations $\delta$ lies in the range $1/3 \le \delta \le 1$
\cite{CH}. Then, the mass of produced PBHs $M_{BH}$ is roughly given
by the horizon mass,
\beq
  M_{BH} \simeq \frac{4 \sqrt{3} \pi}{\sqrt{\rho}}
         \simeq 0.066 M_{\odot} \lmk \frac{T}{\rm GeV} \rmk^{-2},
  \label{eq:BHmass}
\eeq
where $\rho$ and $T$ are the total energy density and the temperature
of the universe at formation. The horizon scale at formation is
related to the present cosmological scale $L$ by
\bea
  L &\simeq& \frac{a(T_{0})}{a(T)} H^{-1}(T), \non \\
    &\simeq& 6.4 \times 10^{-8} {\rm Mpc} 
       \lmk \frac{T}{\rm GeV} \rmk^{-1},\
  \label{eq:BHscale}
\eea
with $T_{0} \simeq 2.7 $ K the present temperature of the universe.
The corresponding comoving wave number $k = 2\pi / L$ is given by
\beq
  k \simeq 1.0 \times 10^{8} {\rm ~Mpc}^{-1} 
                       \lmk \frac{T}{\rm GeV} \rmk.\
  \label{eq:BHwave}
\eeq

Assuming Gaussian fluctuations, the mass fraction of produced PBHs
($\beta \equiv \rho_{BH}/\rho$) is given by
\bea
  \delta(M) &=& \int^{1}_{1/3} \frac{1}{\sqrt{2\pi}\sigma(M)}
               \exp \lmk - \frac{\delta^{2}}{2 \sigma^{2}(M)} \rmk
                d\delta, \non \\
            &\simeq& \sigma(M) 
               \exp \lmk - \frac{1}{18 \sigma^{2}(M)} \rmk,
  \label{eq:BHfraction}
\eea
where $\sigma(M)$ is the root mean square of mass variance evaluated
at horizon crossing. Bullock and Primack \cite{BP} pointed out that
the standard Gaussian assumptions may be inadequate because PBHs are
formed at high density peaks so that the linear theory may be
invalidated and non-Gaussianity may affect the abundance of PBHs
significantly (see also \cite{Ivanov}). However, for example, in the
model adopted in \cite{Ivanov}, the root mean square of mass variance
calculated in the standard Gaussian theory differs at most by the
factor 1.5 from that calculated taking into account the non-Gaussian
effects. The aim in this paper is just to demonstrate that our double
inflation model straightforwardly leads to PBHs formation, and the
concrete values of the parameters should not be taken seriously.
Therefore, for our purpose, it is sufficient to assume that
fluctuations are Gaussian distributed. The analysis of non-Gaussianity
in a similar model as ours is done in Ref. \cite{PBH2}.

Using the mass fraction $\beta$ at formation, the ratio of the present
energy density $\rho_{BH}(M)$ of PBHs with the mass $M$ and the
entropy density is given by
\beq
  \frac{\rho_{BH}(M)}{s} \simeq \frac34 \beta(M) T,
\eeq
which yields the normalized energy density,
\bea
  \Omega_{BH} h^{-2} &\simeq& 2.1 \times 10^{8} \beta 
                           \lmk \frac{T}{\rm GeV} \rmk \non \\   
                     &\simeq& 5.4 \times 10^{7} \beta 
                           \lmk \frac{M}{M_{\odot}} \rmk^{-1/2} \non \\   
                     &\simeq& 2.1 \beta 
                           \lmk \frac{k}{\rm Mpc^{-1}} \rmk.
\eea

Then, MACHO PBHs with mass $\sim 0.1M_{\odot}$ are produced at the
temperature given by
\beq
  T \simeq 0.81 {\rm ~GeV},
\eeq
which corresponds to
\bea
  L &\simeq& 7.9 \times 10^{-8} {\rm ~Mpc}, \non \\
  k &\simeq& 8.1 \times 10^{7} {\rm ~Mpc^{-1}}.
  \label{eq:MACHOscale}
\eea
As easily seen from Eq. (\ref{eq:gpotentialn}), the fluctuations with
the largest amplitude are produced at the onset of new inflation,
which we identify with the formation time of PBHs. As shown later, the
spectrum is so steep that the formation of the PBHs with smaller
masses is suppressed strongly. From Eq. (\ref{eq:MACHOscale}) we
obtain $N'_{new} \simeq 37$, which corresponds to
\beq
  N_{new} \simeq 37 - \ln 92 g.
  \label{eq:MACHON}
\eeq
On the other hand, the present energy density $\Omega_{BH} h^{-2} \sim
0.25$ is explained if the mass fraction is given by
\beq
  \beta \simeq 1.5 \times 10^{-9},
\eeq
which implies the mass variance $\sigma \simeq 0.056$ under the
Gaussian approximation, corresponding to
\beq
  \Phi_{A} \sim 0.04.
  \label{eq:MACHOP}
\eeq
Taking into account Eqs. (\ref{eq:MACHON}), (\ref{eq:MACHOP}), and the
COBE normalization (\ref{eq:COBEnor}), MACHO PBHs are produced in this
model if we take the parameters given by
\bea
  \lambda &\sim& 1.3 \times 10^{-6}, \non \\
  v &\sim& 1.1 \times 10^{-6}, \non \\
  v_{2} &\sim& 0.93 \times 10^{-6},
\eea 
with $g = \lambda / v \simeq 1.2$.

As another example, let us consider PBHs responsible for antiproton
fluxes observed by the BESS experiments \cite{BESS} or short gamma ray
bursts \cite{gamma}. Such PBHs are evaporating now, which leads to the
initial mass $M \sim 3 \times 10^{-19} M_{\odot}$. Then, the
temperature at formation is given by $T \simeq 4.7 \times 10^{8}$ GeV
corresponding to $L \simeq 1.4 \times 10^{-16}$ Mpc and $k \simeq 4.7
\times 10^{16}$ Mpc$^{-1}$. On the other hand, the abundance is given
by $\Omega_{BH}h^{2} \simeq 2 \times 10^{-9}$, which implies $\beta
\simeq 2.0 \times 10^{-26}$, $\sigma \simeq 0.032$, and $\Phi_{A}
\simeq 0.002$. The PBHs satisfying the above conditions are produced
if we take the parameters given by
\bea
  \lambda &\sim& 5.8 \times 10^{-7}, \non \\
  v &\sim& 3.9 \times 10^{-7}, \non \\
  v_{2} &\sim& 5.2 \times 10^{-6},
\eea 
with $g = \lambda / v \simeq 1.5$. Note that in both cases, all
parameters are of the same order. Also, the temperatures at formation
are lower than the reheating temperature so our assumption that PBHs
are formed in the radiation dominated universe is justified.

\subsection{Numerical calculations of density fluctuations}

In this section we numerically calculate density fluctuations
produced during double inflation in order to confirm the analytic
results given above and show explicitly that density fluctuations
which lead to PBHs are realized in our model. Our method of numerical
calculations is based on Ref. \cite{num}.

We decompose multiscalar fields $\phi_{i}(\vect{x},t)$ into the
homogeneous mode $\overline{\phi_{i}}(t)$ and fluctuations
$X_{i}(\vect{x},t)$,
\bea
  \phi_{i}(\vect{x},t) &=& \overline{\phi_{i}}(t) + X_{i}(\vect{x},t)
                               \non \\
                       &=& \overline{\phi_{i}}(t) + 
          \int  \frac{d^{3}\vect{k}}{(2\pi)^{3/2}}
                \lmk
                 \widetilde{X}_{i}(\vect{k},t)
                   e^{i\vect{k}\cdot\vect{x}} +
                   \widetilde{X}^{\dagger}_{i}(\vect{k},t)
                    e^{-i\vect{k}\cdot\vect{x}}
                \rmk.
\eea
The metric is also expanded around the background metric,
\bea
  g_{\mu\nu}(\vect{x},t) &=& \overline{g}_{\mu\nu}(t) 
                              + y_{\mu\nu}(\vect{x},t)  \non \\
             &=& \overline{g}_{\mu\nu}(t) +
          \int  \frac{d^{3}\vect{k}}{(2\pi)^{3/2}}
                \lmk 
                 \widetilde{y}_{\mu\nu}(\vect{k},t)         
                   e^{i\vect{k}\cdot\vect{x}} + 
                 \widetilde{y}^{\dagger}_{\mu\nu}(\vect{k},t)         
                   e^{-i\vect{k}\cdot\vect{x}} 
                \rmk.
\eea
Since fluctuations are generated quantum mechanically, we need to
treat them as quantum Heisenberg operators. Defining creation and
annihilation operators satisfying the canonical commutation relations
$ [\widehat{a}_{i}(\vect{k}),\widehat{a}^{\dagger}_{j}(\vect{k}')]
=\delta_{ij}\delta(\vect{k}-\vect{k}')$, fluctuations are expanded
over such operators:
\bea
  \widetilde{X}_{i}(\vect{k},t) &=&
     \sum_{j} \delta\varphi_{ij}(\vect{k},t) 
                \widehat{a}_{j}(\vect{k}), \non \\
  \widetilde{y}_{\mu\nu}(\vect{k},t) &=& \sum_{i} h_{\mu\nu,i}(\vect{k},t)
                              \widehat{a}_{i}(\vect{k}).
\eea
In particular, the metric perturbation in the longitudinal gauge is
expanded as
\bea
  \Phi_{A}(\vect{x},t) &=& \int \frac{d^{3}\vect{k}}{(2\pi)^{3/2}}
                \lmk 
                   \Phi(\vect{k},t)
                   e^{i\vect{k}\cdot\vect{x}} +
                   \Phi^{\dagger}(\vect{k},t)
                   e^{-i\vect{k}\cdot\vect{x}} 
                \rmk, \\
  \Phi(\vect{k},t) &=& \sum_{i} \Phi_{i}(\vect{k},t)
                              \widehat{a}_{i}(\vect{k}).
\eea

The equations of motion for the homogeneous mode are given by
\beq
  \ddot{\overline{\phi}}_{i} + 3 H \dot{\overline{\phi}}_{i} 
     + \frac{\del V}{\del {\overline{\phi_{i}}}} = 0,
\eeq
with the hubble constant $H$ given by
\beq
  H^{2} = \frac{1}{3M_{G}^{2}}
             \lmk
               V + \frac12 \sum_{i} \dot{\overline{\phi}}^{2}_{i}
             \rmk.
\eeq
Here and hereafter we recover the reduced Planck scale $M_{G}$. The
perturbed equations of motion in the longitudinal gauge are given by
\bea
  \dot{\Phi_{i}} + H \Phi_{i} &=& \frac{1}{2M_{G}^{2}} 
         \sum_{j} 
            \dot{\overline{\phi}}_{i} \delta\varphi_{ij}, 
  \label{eq:dotphi} \\
  \ddot{\delta\varphi}_{ij} + 3 H \dot{\delta\varphi}_{ij} 
       + \lmk \frac{k^{2}}{a^{2}} + 
               \frac{\del^{2}V}{\del{\overline{\phi_{i}}}^{2}}
         \rmk
            \delta\varphi_{ij} 
       &=&
          4 \dot{\overline{\phi}}_{i} \dot\Phi_{j}
          - 2 \frac{\del V}{\del{\overline{\phi_{i}}}} \Phi_{j}
          + \sum_{l \ne i} \frac{\del^{2}V}
               {\del{\overline{\phi_{i}}} \del{\overline{\phi_{l}}}}
                 \delta\varphi_{lj},
  \label{eq:ddotvar}
\eea
for arbitrary $i,j$. There is another convenient equation given by
\beq
  \Phi_{i} = \sum_{j} \lkk \lmk 
               - 3 H \dot{\overline{\phi}}_{j} 
               - \frac{\del V}{\del {\overline{\phi_{j}}}}       
                           \rmk \delta\varphi_{ji}
               - \dot{\overline{\phi}}_{j} \delta\varphi_{ji}
                      \rkk \biggl/
              \lmk 2 M_{G}^{2} \frac{k^{2}}{a^{2}}
                   - \sum_{j} \dot{\overline{\phi}}^{2}_{j} 
              \rmk.
  \label{eq:phi}
\eeq

Before giving the initial conditions for $\delta\varphi_{ij}$ and
$\Phi_{i}$, we determine the normalization of $\delta\varphi_{ij}$.
The conjugate momentum of $X_{i}$ is given by $a^{3}\dot{X}_{i}$,
which leads to the equal-time commutation relation
$[X_{i}(\vect{x},t),\dot{X}_{j}(\vect{x}',t)] = i a^{-3}(t)
\delta_{ij} \delta(\vect{x}-\vect{x}')$. Thus the Wronskian
normalization conditions for $\delta\varphi_{ij}$ are given by
\beq
  \sum_{j} ( \delta\varphi_{ij} \dot{\delta\varphi}_{jl}^{\ast}
            - \delta\varphi_{ij}^{\ast} \dot{\delta\varphi}_{jl} )
          = i a^{-3} \delta_{il}.
\eeq
Then, we find the WKB solutions of Eq. (\ref{eq:ddotvar}) in the short
wavelength approximation ($k/a \gg H$),
\beq
  \delta\varphi_{ij} = \delta_{ij} \frac{1}{\sqrt{2k} a}
                         \exp \lmk -i k \int \frac{dt}{a} \rmk,
\eeq
which give the initial conditions for $\delta\varphi_{ij}$.
Differentiating these with respect to the cosmic time, we obtain the
initial conditions for $\dot{\delta\varphi}_{ij}$,
\beq
  \dot{\delta\varphi}_{ij} = \delta_{ij} 
          \lmk
            - i \sqrt{\frac{k}{2}} \frac{1}{a^{2}}
            - \frac{\dot{a}}{\sqrt{2k} a^{2}}
          \rmk
            \exp \lmk -i k \int \frac{dt}{a} \rmk.
\eeq
Note that the exponent can be set to zero because the origin of the
conformal time $\equiv \int dt/a$ is arbitrary. The initial conditions
for $\Phi_{i}$ are derived from Eq. (\ref{eq:phi}) by the use of those
of $\delta\varphi_{ij}$ and $\dot{\delta\varphi}_{ij}$.

Since the vacuum expectation value of the squared of the operator
$\Phi_{A}(\vect{x},t)$ is given by
\beq
  \la 0 | \Phi_{A}^{2}(\vect{x},t) | 0 \ra =
        \int  \frac{d^{3}\vect{k}}{(2\pi)^{3/2}}
                \sum_{i} |\Phi_{i}(\vect{k},t)|^{2},
\eeq    
we define the primordial spectrum $k^{3/2} \Phi(\vect{k})$
as\footnote{Note that the relation between $k^{3/2} \Phi(\vect{k})$
and $\Phi_{A}(N)$ used in the previous sections is given by
$k^{3/2} \Phi(\vect{k}) \simeq \pi\sqrt{2} \Phi_{A}(N)$.}
\beq
  k^{3/2} \sqrt{\la 0 | \Phi^{2}(\vect{k}) | 0 \ra / \delta(\vect{0})}
    = k^{3/2} \sqrt{\sum_{i} |\Phi_{i}(\vect{k})|^{2}}.
\eeq

All quantities are normalized by the combination of $M_{G}$ and $H_{0}
\equiv v / \sqrt{3}$. Concretely, $\widetilde{\phi}_{i} =
\overline{\phi}_{i} / M_{G}$, $\widetilde{V} = V /
(M_{G}^{2}H_{0}^{2})$, $\widetilde{t} = t H_{0}$,
$\widetilde{\delta\varphi}_{ij} = \delta\varphi_{ij} H_{0}^{1/2}$,
$\widetilde{\Phi}_{i} = \Phi_{i} M_{G} H_{0}^{1/2}$, and
$\widetilde{k} = k H_{0}^{-1}$. Thus all terms except $k^{2}/a^{2}$ in
the equations of motion become of the order of unity, which is
essentially important for numerical calculations, avoiding rounded
errors. Also, in order to confirm the results of our numerical
calculations we compare the spectrum derived from the evolution
equation (\ref{eq:dotphi}) and that obtained from the constrained
equation (\ref{eq:phi}). Both spectra coincide up to the order of
$10^{-8}$.

\begin{figure}[htb]
  \begin{center}
    \leavevmode\psfig{figure=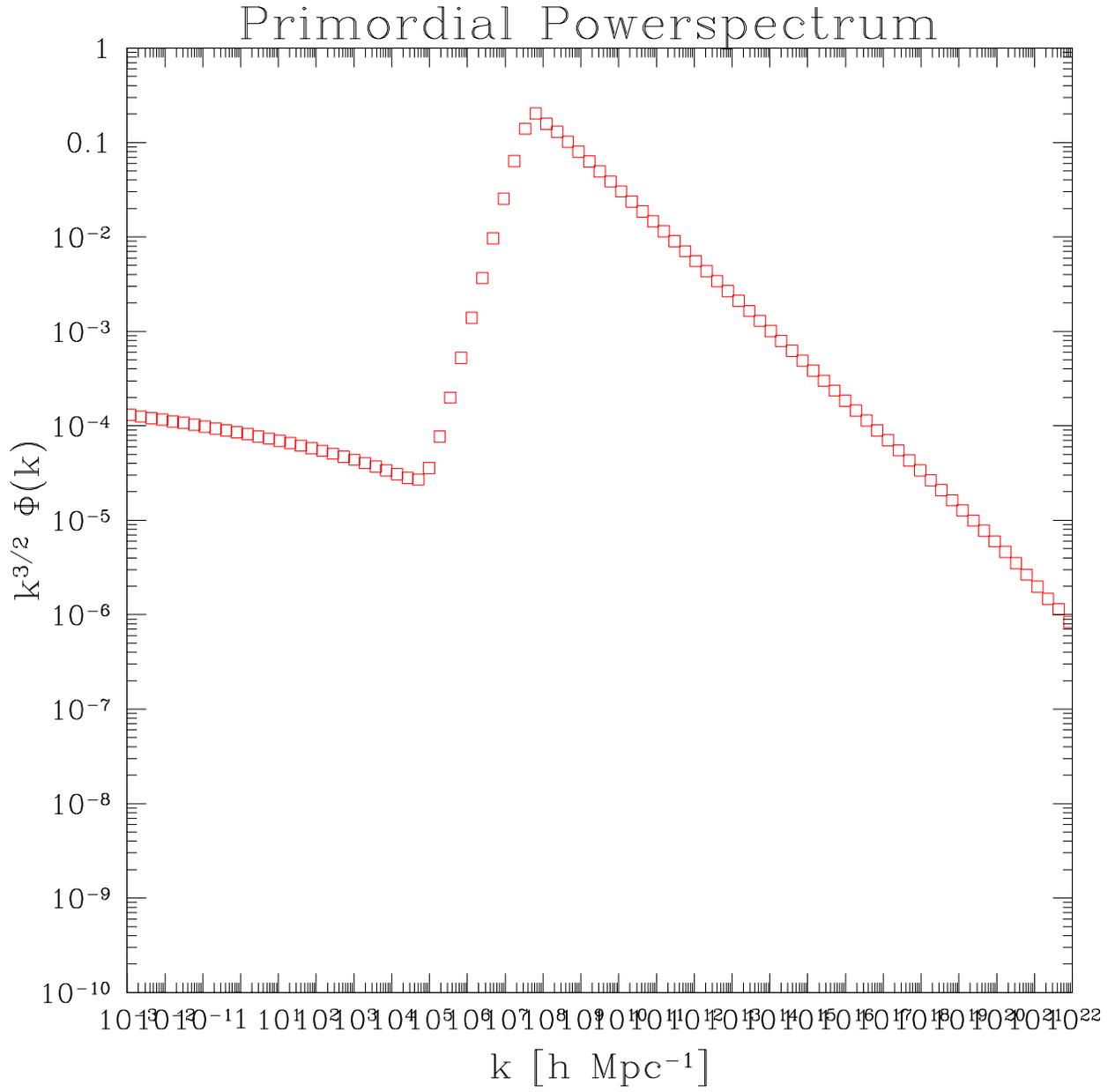,width=17cm}
  \end{center}
  \caption{The spectrum of primordial fluctuations which produces the
  MACHO PBHs is depicted.}
  \label{fig:macho}
\end{figure}

\begin{figure}[htb]
  \begin{center}
    \leavevmode\psfig{figure=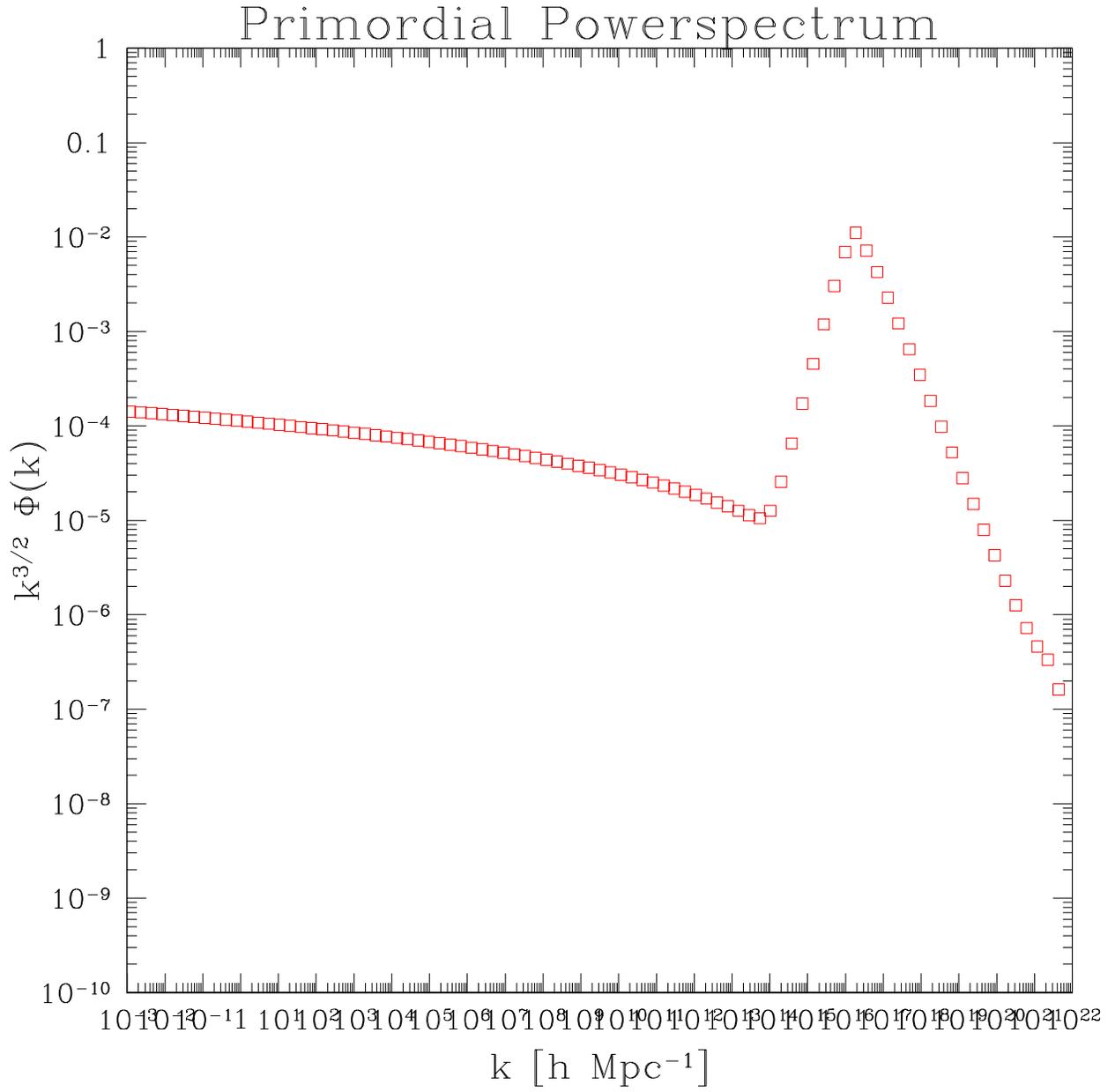,width=17cm}
  \end{center}
  \caption{The spectrum of primordial fluctuations which produces the
  BESS (GRBs) PBHs is depicted.}
  \label{fig:bess}
\end{figure}

The spectra of primordial fluctuations for the cases of the MACHO PBHs
and the BESS (GRBs) PBHs are depicted in Figs. \ref{fig:macho} and
\ref{fig:bess}. As is easily seen, the numerical results reproduce
excellently analytic estimates, which is confirmed to be correct.

\section{Discussion and conclusions}

\label{sec:con}

In this paper we have minutely investigated a natural double
inflation model in SUGRA. By virtue of the shift symmetry, chaotic
inflation can take place, during which the initial value of new
inflation is set. The initial value of new inflation is adequately far
from the local maximum of the potential so that primordial
fluctuations within the present horizon scale are attributed to both
inflations. That is, fluctuations responsible for the anisotropy of
the CMB and the large scale structure are produced during chaotic
inflation, while fluctuations on smaller scale are produced during new
inflation. Due to the peculiar nature of new inflation, fluctuations
on smaller scale are as large as of the order of unity, which may lead
to PBHs formation. As examples we consider PBHs responsible for
MACHOs and antiproton flux observed by the BESS experiment or short
gamma ray bursts. We find that if we take reasonable values of
parameters, such PBHs are produced in our double inflation model. To
make sure, we also perform numerical calculations and confirm analytic
estimates definitely.

\subsection*{Acknowledgments}

M.Y. is grateful to T. Kanazawa, M. Kawasaki, F. Takahashi, and J.
Yokoyama for discussions. M.Y. is partially supported by the Japanese
Society for the Promotion of Science.

\begin{table}[t]
  \begin{center}
    \begin{tabular}{| c | c | c | c | c |}
                   & $\Phi$ & $X$ & $\Xi$ & $\Pi$ \\
        \hline
        $Q_R$      & 0      & 2   & 0     & 0 \\ 
        \hline 
        $Z_{2}$    & $-$    & $-$ & $-$   & $-$   
    \end{tabular}
    \caption{The charges of various supermultiplets of U$(1)_{R}
    \times Z_{2}$.} 
    \label{tab:charges}
  \end{center}
\end{table}

\end{document}